\RequirePackage{pdf14}

\documentclass[letterpaper]{article}
\usepackage[usenames,dvipsnames,table]{xcolor}
\usepackage{graphicx}
\usepackage{caption}
\usepackage{subcaption}
\usepackage{array,setspace,mathrsfs,amsthm,amsfonts,amsmath,yfonts,dsfont,bbm,colonequals,mathtools}
\usepackage{./aux/jheppub}
\usepackage{afterpage}
\usepackage{xspace}
\usepackage[normalem]{ulem}

\newcommand\be{\begin{equation}}
\newcommand\bea{\begin{eqnarray}}
\newcommand\eea{\end{eqnarray}}
\newcommand\ee{\end{equation}}

\newcommand\eg{{\it e.g.}}
\newcommand\ie{{\it i.e.}}

\makeatletter
\def\maketag@@@#1{\hbox{\m@th\normalfont\normalsize#1}}
\makeatother

\def\BB{{\mathcal{B}}}
\def\CC{{\mathcal{C}}}

\def\NN{{\mathcal{N}}}

\def\Om{{\mathcal{O}}}
\def\Cm{{\mathcal{C}}}

\def\Bm{{\mathcal{B}}}
\def\Nm{{\mathcal{N}}}

\def\Dm{{\mathcal{D}}}
\def\Qm{{\mathcal{Q}}}

\def\pd{\partial}

\def\nn{\nonumber}

\newcommand\SL{\mathfrak{sl}}
\newcommand\SO{\mathfrak{so}}
\newcommand\SU{\mathfrak{su}}
\newcommand\USp{\mathfrak{usp}}
\newcommand\E{\mathfrak{e}}
\newcommand\F{\mathfrak{f}}
\newcommand\G{\mathfrak{g}}

\title{$\Nm=2$ central charge bounds from $2d$ chiral algebras}
\author[1]{Madalena Lemos,}
\author[2]{Pedro Liendo}

\affiliation[1]{DESY Hamburg, Theory Group, Notkestrasse 85, D–22607 Hamburg, Germany}
\affiliation[2]{IMIP, Humboldt-Universit{\"a}t zu Berlin, IRIS Adlershof, Zum Gro{\ss}en Windkanal 6, 12489 Berlin, Germany}

\emailAdd{madalena.lemos@desy.de}
\emailAdd{pliendo@physik.hu-berlin.de}

\preprint{DESY 15-230, HU-EP-15/56}

\bigskip
\abstract{
We study protected correlation functions in $\Nm = 2$ SCFT whose description is captured by a two-dimensional chiral algebra. Our analysis implies a new analytic bound for the $c$-anomaly as a function of the flavor central charge $k$, valid for any theory with a flavor symmetry $G$. Combining our result with older bounds in the literature puts strong constraints on the parameter space of $\Nm=2$ theories. In particular, it singles out a special set of models whose value of $c$ is uniquely fixed once $k$ is given. This set includes the canonical rank one $\Nm=2$ SCFTs given by Kodaira's classification.
}


\begin{document}
\setcounter{tocdepth}{1}
\maketitle
\setcounter{page}{1}

\section{Introduction and summary}
\label{sec:intro}

Every four-dimensional $\NN=2$ superconformal field theory (SCFT) contains a protected subsector isomorphic to a two-dimensional chiral algebra \cite{Beem:2013sza}. This observation has led to interesting insights into the structure of $\NN=2$ superconformal dynamics, and it will be our main tool in this work. The construction starts by considering a certain class of operators that belong to the cohomolgy of a nilpotent supercharge. These operators are restricted to lie on a plane, and their correlation functions are a meromorphic function of the coordinates on the plane.
Two-dimensional chiral algebras are very rigid structures, allowing to fix an infinite amount of protected data from the parent four-dimensional theory.
To name a few, it allows to fix selected correlation functions, predict Higgs branch relations, and put constraints on central charges \cite{Beem:2014rza,Lemos:2014lua,Liendo:2015ofa}.
There has also been a very fruitful interplay between these chiral algebras and the superconformal index \cite{Chacaltana:2014nya,Buican:2015ina,Cordova:2015nma,Buican:2015tda,Song:2015wta,Cecotti:2015lab}.

A similar construction was considered in six-dimensions, allowing the computation of certain exact three-point functions \cite{Beem:2014kka}; and in three dimensions \cite{Chester:2014mea}, where the cohomological argument implies a subset of operators is described by a one-dimensional topological theory.

In this note we will be interested in central charge bounds, which can be obtained thanks to a subtle interplay between two-dimensional and four-dimensional quantities. As will be reviewed below, the existence of bounds is closely connected to the presence of null states in the chiral algebra. 

Our assumptions will be minimal: an interacting theory with a unique stress-tensor, and a flavor current generating a global flavor symmetry. These two conserved currents sit in shortened multiplets of the $\NN=2$ superconformal algebra, that contain an element captured by the chiral algebra. These multiplets have been studied individually from the chiral perspective in \cite{Beem:2013sza} and \cite{Liendo:2015ofa}, leading to new unitarity bounds.\footnote{Similarly, a unitarity bound for $\NN=4$ SCFTs was obtained in \cite{Beem:2013qxa}.} Here we consider the logical next step and study the most natural mixed system of correlators: all possible combinations between flavor and stress-tensor multiplets. This analysis will lead to a new analytic bound that when complemented with older results severely reduces the parameter space of $\NN=2$ theories. This bound constrains, for each flavor group, the allowed values of the flavor central charge $k_{4d}$ and the $c_{4d}$-anomaly coefficient:
\be
k_{4d} \left(-180 c_{4d}^2+66 c_{4d}+3 \mathrm{dim}_G\right)+60 c_{4d}^2 h^\vee-22 c_{4d} h^\vee \leqslant 0\,,
\label{eq:newbound_intro}
\ee
where the group information enters through the dual Coxeter number $h^\vee$ and the dimension of of the algebra $\mathrm{dim}_G$.

This bound carves out the low $c_{4d}$ portion of the $(c_{4d},k_{4d})$ plane, explaining why there are no known theories in this region of the $\Nm=2$ landscape. This new bound does not only give a general constraint valid for any theory, but it will also allow us to learn more about a specific set of models that sit at special corners of the allowed parameter space. 
\begin{table}
\centering
\renewcommand{\arraystretch}{1.3}
\begin{tabular}{|c||c|c|c|c|c|c|c|}
\hline
$G$	&	 $H_1$ 	       & $H_2$ 	        & $D_4$ 	      & $E_6$ 		    & $E_7$ 		  & $E_8$ 			\\ \hline\hline 
$c_{4d}$	 	& $\frac{1}{2}$  & $\frac{2}{3}$  & $\frac{7}{6}$   & $\frac{13}{6}$  & $\frac{19}{6}$  & $\frac{31}{6}$  \\ \hline
$k_{4d}$		& $\frac{8}{3}$  	& $3$ 		    & $4$ 		      & $6$ 			& $8$ 			  & $12$ 			\\ 
\hline
\end{tabular}
\caption{Properties of the canonical rank one SCFTs associated to maximal mass deformations of the Kodaira singularities with a flavor symmetry \cite{Argyres:1995xn,Minahan:1996fg,Minahan:1996cj,Cheung:1997id,Argyres:2007cn,Aharony:2007dj}. The flavor symmetry of the theories is given by $G$ with $H_i \to A_i$. They can also be obtained as the low energy theory living on a single $D3$-brane probing 
an $F$-theory singularity of type $G$
\cite{Sen:1996vd,Banks:1996nj,Dasgupta:1996ij,Minahan:1996fg,Minahan:1996cj,Aharony:1998xz}.
\label{Tab:rank1theories}}
\end{table}

We present them in Tab.~\ref{Tab:rank1theories} along with their values of $c_{4d}$ and $k_{4d}$, and a short description on ways in which they can be obtained. These models all share the property that their Higgs branch is given by the one-instanton moduli space of $G$, with their Higgs branch chiral rings finitely generated by the moment maps subject to a relation defining the ``Joseph ideal'' (see \eg, \cite{Gaiotto:2008nz}). In fact, in  \cite{Beem:2013sza}, the Higgs branch relation was formulated as a saturation of a unitarity bound, and it was shown that only a restricted number of flavor symmetries can satisfy this constraint: $G=A_1,A_2,D_4,E_6,E_7,E_8,G_2,F_4$.
Moreover, for this to happen the values of $c_{4d}$ and $k_{4d}$ are fixed to specific values (the only exception being  $A_1$\footnote{For generic $G$ the relations obtained implied two constraints that fix $c_{4d}$ and $k_{4d}$, while for $A_1$ there is only one constraint that fixes $c_{4d}$ as a function of $k_{4d}$.}). 
These values matched precisely the ones given in Tab.~\ref{Tab:rank1theories}, and predicted two more cases which do not correspond to known theories: $G_2$ and $F_4$. This was an elegant way of rederiving Tab.~\ref{Tab:rank1theories}, and one can now ask the interesting question of how unique these theories are. Namely, if one considers a generic theory assuming nothing about its Higgs branch, and we fix $k_{4d}$ to one of the values given in Tab.~\ref{Tab:rank1theories}, what can we say about the value of $c_{4d}$? The results of \cite{Beem:2013sza} allow for a range of $c_{4d}$ values for each of the $k_{4d}$ in the table, and the question of uniqueness remained unanswered.

The new analytic bound obtained here \eqref{eq:newbound_intro}, in combination with the old bounds of \cite{Beem:2013sza}, will succeed in singling out the canonical set of Tab.~\ref{Tab:rank1theories} with minimal physical input. We will see that these theories live at the intersection of the old and new bound: given $k_{4d}$ from Tab.~\ref{Tab:rank1theories}, the old bound acts as an upper bound on the central charge $c_{4d}$, while the new becomes a lower bound on $c_{4d}$, allowing for a unique value. The Higgs branch relation then follows as a consequence!

The new bound \eqref{eq:newbound_intro} also has implications for the numerical bootstrap program for $\Nm=2$ theories initiated in \cite{Beem:2014zpa,Lemos:2015awa}. It indicates that the set of correlators studied in this paper is a promising way to bootstrap the canonical set of Tab.~\ref{Tab:rank1theories}, along with the mysterious $G_2$ and $F_4$ cases. While $D_4$ is the familiar $\SU(2)$ superconformal QCD, the remaining ones have no known Lagrangian description, and the numerical bootstrap seems to be the best approach in order to access their non-protected sector. The bootstrap program for $\Nm=2$ theories 
can be thought of as a two-step process. The first one consists in fixing the part of the correlators captured by the chiral algebra, and obtaining from them an infinite number of OPE coefficients. The second step involves the numerical bootstrap of the complete system of correlators, constraining the CFT data not captured by the chiral algebra. For the second step it is necessary to write the superconformal block decomposition of these correlators. While the blocks for the superconformal primary of the flavor current multiplet four-point function have been obtained in the literature \cite{Dolan:2001tt,Nirschl:2004pa,Dolan:2004mu}, the remaining ones are still unknown. Still, the results of this paper fulfil the first requirement, and provide the starting point toward the full-blown numerical bootstrap for this system of correlators.
\section{A new analytic bound}
\label{sec:fourpt}

Let us start by reviewing the $2d$ chiral algebra description of $4d$ $\Nm=2$ SCFTs. We refer the reader to \cite{Beem:2013sza} for more details. The construction relies on the existence of an $\SU(1,1|2)$ subalgebra of the full superconformal algebra with an $\overline{\SL}(2) \times \SU(2)_R$ factor, and a nilpotent supercharge $\mathbbmtt{Q}$. It can be shown that the diagonal subalgebra of $\overline{\SL}(2) \times \SU(2)_R$, denoted by $\widehat{\SL(2)}$, is $\mathbbmtt{Q}$-exact.

We restrict local operators to a plane with coordinates $(z,\bar{z})$, of which $\overline{\SL}(2)\times \SL(2)$ is the global conformal algebra. If an operator at the origin is in the cohomology of $\mathbbmtt{Q}$, we can use the twisted $\widehat{\SL(2)}$ to generate anti-holomorphic translations such that it remains in the cohomology. Moreover, the anti-holomorphic dependence of the twisted-translated operator will be $\mathbbmtt{Q}$-exact, and it is then an easy exercise to prove that it will cancel from correlation functions. What remains is just a meromorphic correlator:
\be 
\langle \Om(z_1,\bar{z}_1)\Om(z_2,\bar{z}_2)\Om(z_3,\bar{z}_3)\Om(z_4,\bar{z}_4) \rangle = f(z_1,z_2,z_3,z_4)\, . 
\ee
The cohomology of $\mathbbmtt{Q}$ is non-empty and its cohomology classes are in one-to-one correspondence with certain short multiplets of the $\Nm=2$ superconformal algebra. We call the operators described by the $2d$ chiral algebra ``Schur operators'', because they sit in multiplets that contribute to the Schur limit of the superconformal index \cite{Kinney:2005ej,Gadde:2011ik,Gadde:2011uv}. In Tab.~\ref{schurTable} we list all the multiplets that contain Schur operators and their $4d$ and $2d$ quantum numbers.
\renewcommand{\arraystretch}{1.5}
\begin{table}
\centering
\begin{tabular}{|l|l|l|l|l|}
\hline
Multiplet  & $(\Delta,j,\bar{\jmath},R,r)$ & $\Om_{\rm Schur}$  & $h$ & $r$    \\ 
\hline 
$\hat \Bm_R$  & $(2R,0,0,0,R,0)$ &  $\Psi^{11\dots 1}$   &    $R$ &  $0$ \\ 
\hline
$\Dm_{R (0, \bar{\jmath})}$  & $(2R + \bar{\jmath} +1,0,\bar{\jmath},R,\bar{\jmath} +1)$ &    $ \bar{\Qm}^1_{\dot{+}} \Psi^{11\dots 1}_{\dot  + \dots \dot  + }$ &   $R+ \bar{\jmath} +1$  & $\bar{\jmath} + \frac{1}{2}$  \\
\hline
$\bar \Dm_{R (j, 0 )}$  & $(2R + j +1,j,0,R,-j-1)$ & $ {\Qm}^1_{ +} \Psi^{11\dots 1}_{+   \dots +}$ &     $R+ j+1$  & $-j - \frac{1}{2}$  \\
\hline
$\hat \Cm_{R (j, \bar{\jmath}) }$ & $(2R + j + \bar{\jmath} + 2,j,\bar{\jmath},R,0)$ &   ${\Qm}^1_{+} \bar{\Qm}^1_{\dot{+}} \Psi^{11\dots 1}_{+   \dots + \, \dot  + \dots \dot  + }$&   
 $R+ j + \bar{\jmath} +2$ &  $\bar{\jmath} - j$  \\
\hline
\end{tabular}
\caption{\label{schurTable} Short multiplets of that contain Schur operators using the notation of \cite{Dolan:2002zh}. The second column gives the associated Dynkin labels: $\Delta$ corresponds to dilatations, $(j,\bar{\jmath})$ label Lorentz representations, and $(R,r)$ are the $\SU(2)_R \times \mathfrak{u}(1)_r$ quantum numbers. The third column indicates where inside the multiplet the Schur operator sits, where $(+,\dot{+},1)$ are the highest-weight values for $(j,\bar{\jmath},R)$. The last two columns give the $2d$ holomorphic dimension and $r$-charge in terms of $(j,\bar{\jmath},R)$.
} 
\end{table}
The two multiplets of interest to us are
\begin{itemize}

\item $\hat{\Cm}_{0(0,0)}$, conserved semi-short multiplet that contains the stress-tensor and the conserved $\SU(2)_R$ and $\mathfrak{u}(1)_r$ currents.

\item $\hat{\Bm}_{1}$, conserved short multiplet that contains the flavor current, and therefore transforms in the adjoint of the flavor group. Its highest weight is the moment map operator: a scalar of dimension $\Delta=2$ that transforms in the triplet of $\SU(2)_R$.
\end{itemize}
We will be interested in the four-point functions of the Schur operators of these multiplets, as they are the ones described by the chiral algebra. For the $\hat{\Cm}_{0(0,0)}$ multiplet this corresponds to the $\SU(2)_R$ current: $J_{+\dot{+}}^{IJ}$, and we denote the associated $2d$ operator by $T$. For the $\hat{\BB}_1$ multiplet the Schur operator is the moment map and we denote the corresponding $2d$ operator by $J^a$, with $a$ an adjoint index.
As shown in \cite{Beem:2013sza}, the $4d$ OPEs imply the following $2d$ OPEs:
\begin{align}
\begin{split}
T(z)T(0) & \sim -\frac{6\,c_{4d}}{z^4} + 2 \frac{T(0)}{z^2} + \frac{\pd T(0)}{z} + \ldots\, ,
\\
J^a(z)J^b(0) & \sim -\frac{k_{4d}}{2}\frac{ \delta^{ab}}{z^2} + i f^{abc}\frac{J^c(0)}{z} + \ldots\, .
\end{split}
\end{align}
The global $\SL(2)$ enhances to full Virasoro and $T(z)$ becomes the $2d$ stress-tensor. Similarly the global flavor symmetry gives rise to the the holomorphic current $J^a(z)$ of an affine Kac-Moody algebra at level $k_{2d}$. The $2d$ central charges $c_{2d}$ and $k_{2d}$ are related to their four-dimensional counterparts by
\be 
\label{c2dc4d}
c_{2d} = -12\, c_{4d}\, , \qquad k_{2d} = -\frac{1}{2}k_{4d}\,.
\ee

\subsection{Schur correlators in \texorpdfstring{$4d$}{4d}}

In this work we are only interested in correlators of Schur operators because they are captured by the $2d$ chiral algebra, and not in the full content of the $\hat{\CC}_{0,(0,0)}$ and $\hat{\BB}_1$ multiplets. In particular we will consider the following mixed system of correlators:
\begin{align}
\label{Schurcorrelators}
\langle \hat{\BB}_1  \hat{\BB}_1  \hat{\BB}_1  \hat{\BB}_1 \rangle\,,\quad \langle \hat{\CC}_{0,(0,0)}\hat{\CC}_{0,(0,0)}\hat{\CC}_{0,(0,0)}\hat{\CC}_{0,(0,0)} \rangle\,,\quad \langle \hat{\CC}_{0,(0,0)}\hat{\CC}_{0,(0,0)}  \hat{\BB}_1  \hat{\BB}_1 \rangle\, .
\end{align}
The OPE selection rules relevant for this set of correlators are,
\begin{align}
\begin{split}
\label{SchurOPEexpansions}
\hat{\BB}_1 \times \hat{\BB}_1 &\sim \mathbf{1}+\hat{\BB}_1+\hat{\BB}_2+\hat{\CC}_{0(j,j)}+\hat{\CC}_{1(j,j)}\,,\\
\hat{\CC}_{0,(0,0)} \times \hat{\CC}_{0,(0,0)} &\sim \mathbf{1}+\hat{\CC}_{0(j,j)}+\hat{\CC}_{1(j,j)}\,,\\
\hat{\CC}_{0,(0,0)} \times \hat{\BB}_1 &\sim \hat{\BB}_1+\hat{\BB}_2+\hat{\CC}_{0(j,j)}+ \hat{\CC}_{1(j,j)}\,.
\end{split}
\end{align}
Here we are only listing the exchange of Schur operators.\footnote{See \cite{Arutyunov:2001qw,Liendo:2015ofa} for the full OPE selection rules in the $\hat{\BB}_1 \times \hat{\BB}_1 $ and $\hat{\CC}_{0,(0,0)} \times \hat{\CC}_{0,(0,0)}$ channels. The $\hat{\CC}_{0,(0,0)} \times \hat{\BB}_1$ channel will appear in \cite{Israel}.} By an abuse of notation we denote the Schur operator coming from a given multiplet by the name of the corresponding multiplet.
For the last OPE we only used $\SU(2)_R$ and $\mathfrak{u}(1)_r$ selection rules, which are necessary but not sufficient conditions for multiplets to appear.
The multiplets $\hat{\CC}_{0(j,j)}$ with $j > 0$ contain conserved currents of spin larger than two, and therefore they are absent in an interacting theory \cite{Maldacena:2011jn,Alba:2013yda}.
In the last OPE all operators must transform in the adjoint of the flavor group, and since we are interested in interacting theories the $\hat{\CC}_{0(0,0)}$ multiplet will be absent as well, as we expect a unique stress-tensor in the singlet representation. 

A subtle point that plays a key role in our analysis, is that the selection rules cannot distinguish between several copies of the same multiplet appearing in \eqref{SchurOPEexpansions}. In some cases when there is extra physical information one can specify their number. For example, we know there is a single stress-tensor multiplet $\hat{\CC}_{0,(0,0)}$, and that in the $\hat{\BB}_1 \times \hat{\BB}_1$ OPE only the same $\hat{\BB}_1$ appears. Generically, however, in order to distinguish between copies one would need a careful analysis of all possible correlators of the theory. Here we will be content with the mixed system presented in \eqref{Schurcorrelators}. We can split the operators between two groups: the ones appearing in the $\hat{\CC}_{0,(0,0)} \times \hat{\CC}_{0,(0,0)} $ OPE, and the ones appearing in the $\hat{\BB}_1 \times \hat{\BB}_1$ OPE. When considering the first two four-point functions, only operators in either one of the groups are exchanged, but when considering the last four-point function operators in the intersection of the two groups are exchanged.
We will be particularly interested in the $\hat{\Cm}_{1(\frac12,\frac12)}$ multiplet, because this is the one that will give the strongest bound.
\paragraph{An orthonormal basis:} Without loss of generality, let us pick an orthonormal basis of $\hat{\Cm}_{1(\frac12,\frac12)}$ such that only one multiplet appears in the $\hat{\CC}_{0,(0,0)} \times \hat{\CC}_{0,(0,0)} $ OPE, and denote this multiplet by $\hat{\Cm}^{(1)}_{1(\frac12,\frac12)}$, then,
\be 
\hat{\CC}_{0,(0,0)} \times \hat{\CC}_{0,(0,0)} \sim \ldots  + \hat{\Cm}^{(1)}_{1(\frac12,\frac12)}+ \ldots \,.
\ee
This is possible as long as $\langle \hat{\CC}_{0,(0,0)}\hat{\CC}_{0,(0,0)} \hat{\CC}_{1(\frac12,\frac12)} \rangle \neq 0$, which is always the case for $c > \tfrac{11}{30}$ \cite{Liendo:2015ofa}. 
Using this basis, the second group of operators appearing in the $\hat{\BB}_1  \hat{\BB}_1$ OPE contain  $\hat{\Cm}^{(1)}_{1(\frac12,\frac12)}$, plus extra multiplets which we collectively denote by $\hat{\CC}^{(2)}_{1(\frac12,\frac12)}$:
\be 
\hat{\BB}_1 \times  \hat{\BB}_1 \sim \ldots + \hat{\Cm}^{(1)}_{1(\frac12,\frac12)}+ \hat{\CC}^{(2)}_{1(\frac12,\frac12)} + \ldots \,.
\ee
Note that if we consider additional four-point functions we would be able to distinguish the different multiplets contained in $\hat{\CC}^{(2)}_{1(\frac12,\frac12)}$. From our point of view however, we can treat this collection of multiplets as a single entity, since orthogonality implies it contributes to the block expansion as a sum of squares of OPE coefficients, which is also a non-negative number. 
Each of the $\hat{\CC}^{(2)}_{1(\frac12,\frac12)}$ multiplets gives rise to a two-dimensional chiral operator, contributing to the twisted four-point functions as quasi-primary operators with holomorphic dimension $h=4$ (see Tab.~\ref{schurTable}). 
\subsection{Twisted correlators in \texorpdfstring{$2d$}{2d}}

Let us now consider the two-dimensional system of correlators:
\begin{align}
\langle T(z_1) T (z_2) T (z_3) T(z_4) \rangle &  = \frac{1}{z_{12}^4z_{34}^4}G_{TTTT}(z)\,, \\
\label{eq:TTJJcorr}
\langle T(z_1) T (z_2) J^a (z_3) J^b(z_4) \rangle & = \frac{1}{z_{12}^4z_{34}^2}G^{ab}_{TTJJ}(z)\,,\\
\langle J^a(z_1) J^b (z_2) J^c (z_3) J^d(z_4) \rangle & = \frac{1}{z_{12}^2z_{34}^2}G^{abcd}_{JJJJ}(z)\,.
\end{align}
These correlators can be fixed using crossing symmetry or $2d$ Ward identities (see \eg, \cite{Belavin:1984vu}):
\begin{align}
G_{TTTT}(z) & = 1 + z^4 + \frac{z^4}{(1-z)^4} + \frac{8}{c_{2d}}\left(z^2+z^3 + \frac{z^4}{(1-z)^2}+\frac{z^4}{1-z}\right)\, ,
\\
\label{TTJJ}
G^{ab}_{TTJJ}(z) & = \delta^{ab}\frac{c_{2d} (1 - z)^2 + 2 z^2 (2 - 2 z + z^2))}{c_{2d} (1 - z)^2}\, ,
\\
\label{JJJJ}
G^{abcd}_{JJJJ}(z) & =\delta^{ab} \delta^{cd}+z^2\delta^{ac} \delta^{bd}+\frac{z^2}{(1-z)^2} \delta^{ad} \delta^{cb}-\frac{z}{k_{2d}}f^{ace}f^{bde}-\frac{z}{k_{2d}(z-1)}f^{ade}f^{bce}\, .
\end{align}
\begin{table}[tbh] 
\centering
\begin{tabular}{lll||lll}
\hline \hline
$G$ 			&	$h^{\vee}$		& $\mathrm{dim}_{G}$		&	$G$ 			&	$h^{\vee}$		& $\mathrm{dim}_{G}$		\\[0.3ex]
\hline 
$SU(N)$		&	$N$				& $N^2-1$				&	$F_4$		&	$9$				& $52$					\\[0.3ex]
$SO(N)$		& 	$N-2$			& $\frac{N(N-1)}{2}$		&	$E_6$~~~		&	$12$	~~~			& $78$~~~~				\\[0.3ex]
$USp(2N)$	&	$N+1$			& $N(2N+1)$				&	$E_7$		&	$18$				& $133$					\\[0.3ex]
$G_2$		&	$4$				& $14$					&	$E_8$		&	$30$				& $248$					\\[0.3ex]
\hline
\end{tabular} 
\caption{Dual Coxeter number and dimension of the adjoint representation for the simple Lie groups.
\label{Tab:groupinfo}}
\end{table} 

For the last four-point function \eqref{JJJJ} we are only interested in operators being exchanged in the singlet channel, because these are the ones that contribute to the mixed correlator \eqref{TTJJ}. As such, we project $G^{abcd}_{JJJJ}(z)$ onto the singlet and from here on consider only
\be 
G^{abcd}_{JJJJ}(z)\bigg\vert_{\text{singlet}}= \delta^{ab} \delta^{cd}\left(1+\frac{z^2}{{\mathrm{dim}_G}} \left(\frac{2 h^\vee}{k_{2d}(1- z)}+\frac{1}{(z-1)^2}+1\right)\right) \,,
\ee
where $\mathrm{dim}_G$ and $h^\vee$ are the dimension of the algebra and the dual Coxeter number, and are given in Tab.~\ref{Tab:groupinfo} for convenience.
Each correlator is described by the $2d$ chiral algebra and we can therefore expand them using $\SL(2)$ conformal blocks:
\be 
\label{sl2expansion}
G(z) = 1 + \sum_{\substack{h=2}}^{\infty} a^{(h)} g_h(z) \,, \qquad g_h(z) = z \left(-\frac{z}{2}\right)^{h -1} \, _2F_1(h ,h ;2 h ;z)\, , \qquad \text{$h$ even}\, .
\ee
The flavor structure is now trivial so we omit it from the above expansion. The explicit expressions for the $a^{(h)}$ coefficients are 
\begin{align}
\begin{split}
a_{TTTT}^{(h)} =& -\frac{2^{h-2} (h-3) (h-2)^2 (h-1) \frac{h-4}{2}! \left(\frac{h}{2}+1\right)!! \frac{h}{2}!! (h+1)!!}{9 (h-2)!! (2 h-3)!!} \nn \\
&-\frac{2^{\frac{h}{2}+3} \left(h^2-h-1\right) (h-1)!! \Gamma \left(\frac{h}{2}\right)}{c_{2d} (2 h-3)!!} \,,\\
a_{TTJJ}^{(h)} =& -\frac{4 (h-1) h!}{c_{2d} (2 h-3)!!} \,,\\
a_{JJJJ}^{(h)} =& -\frac{2^{h-2} (h-1) ((h-2)!)^2 }{\mathrm{dim}_G (2 (h-2)+1)!} \left(2 (h-1) h+\frac{4 h^\vee}{k_{2d}}\right) \,.
\end{split}
\end{align}
Each $\SL(2)$ block receives contributions from four-dimensional multiplets containing Schur operators, and the interplay between $2d$ and $4d$ perspectives will allow us to obtain bounds.
Let us then interpret the coefficients of the $\SL(2)$ block decomposition in terms of the $2d$ OPE coefficients. From the exchange of $h=4$ operators in the $2d$ correlators we find
\begin{align}
a[T,T,\Om^{(1)}_{h=4}] & = a_{TTTT}^{(h=4)} =\frac{88}{15 c_{4d}}-16\, ,
\\
a[J,J,\Om^{(1)}_{h=4}] + a[J,J,\Om^{(2)}_{h=4}] & = a_{JJJJ}^{(h=4)} =\frac{16 (h^\vee-3 k_{4d})}{5 \mathrm{dim}_G k_{4d}}\, ,
\\
a[T,T,\Om^{(1)}_{h=4}]a[J,J,\Om^{(1)}_{h=4}]  & =  \left(a_{TTJJ}^{(h=4)}\right)^2 = \frac{64}{25 c_{4d}^2}\, ,
\end{align}
which is a system of equations we can solve for $a[J,J,\Om^{(1)}_{h=4}] $ and $ a[J,J,\Om^{(2)}_{h=4}]$.
These two objects determine the OPE coefficients $\lambda^2[\hat{\Bm}_{1},\hat{\Bm}_{1},\hat{\Cm}^{(1)}_{1(\frac12,\frac12)}]$ and $\lambda^2[\hat{\Bm}_{1},\hat{\Bm}_{1},\hat{\Cm}^{(2)}_{1(\frac12,\frac12)}]$ respectively. For that it is necessary work out the exact relation between the $4d$ and the $2d$ OPE coefficients, which corresponds to a theory independent numerical factor. This was done in \cite{Beem:2013sza,Beem:2014zpa} for the $\hat{\BB}_1$ correlator (using the superconformal blocks of \cite{Dolan:2001tt,Nirschl:2004pa,Dolan:2004mu}). Defining $\lambda^2$ as in (3.22) of \cite{Beem:2014zpa}, we can now translate the unitarity constraints in four dimensions to constraints on $a[J,J,\Om^{(1)}_{h=4}] $ and $ a[J,J,\Om^{(2)}_{h=4}]$.
Four-dimensional unitarity requires the $\lambda^2$ to be positive, and from
\be
-\frac{1}{12} a[J,J,\Om^{(2)}_{h=4}] = \lambda^2[\hat{\Bm}_{1},\hat{\Bm}_{1},\hat{\Cm}^{(2)}_{1(\frac12,\frac12)}]  \geqslant 0\,,
\ee
we get a nontrivial constraint on the allowed values of $c_{4d}$ and $k_{2d}$:
\be
\label{eq:newbound}
k_{4d} \left(-180 c_{4d}^2+66 c_{4d}+3 \mathrm{dim}_G\right)+60 c_{4d}^2 h^\vee-22 c_{4d} h^\vee\leqslant 0\,. 
\ee
Note that the only assumptions required to obtain this bound were the existence of a stress tensor, a flavor symmetry $G$, and that the theory is interacting. It provides a universal lower bound for the central charge $c_{4d}$ as a function of $k_{2d}$ valid for any theory with a flavor symmetry.

Before we proceed to a more careful analysis of \eqref{eq:newbound}, let us comment that the same logic can be applied to higher-spin $\hat{\Cm}_{1(j,j)}$ operators, but the associated bounds are weaker. One could also ask whether different bounds could arise from analyzing different OPE channels or more general correlators.   
Decomposing \eqref{eq:TTJJcorr} in $\SL(2)$ blocks by taking the double OPE $\left(T(z_1) J^a(z_2) \right)$ and $\left( J^b(z_3) T(z_4) \right)$ does lead to a $k_{4d}$-independent bound on $c_{4d}$, but it turns out to be weaker than the $c_{4d} \geqslant \tfrac{11}{30}$ bound of \cite{Liendo:2015ofa}. Similarly, from an analysis of the $\langle T(z_1) J^a(z_2) J^b(z_3) J^c(z_4) \rangle$ correlator only weaker bounds on $c_{4d}$ and $k_{4d}$ are obtained. This exhausts all four-point functions of the minimal set given by $T$ and $J$.

One can also enlarge the minimal set by adding extra multiplets. For example, any theory in which the two-dimensional stress tensor is not given by the Sugawara construction will have a $\hat{\Bm}_2$ multiplet associated to the difference $c_{\text{Sug}} T- c\,T_{\text{Sug}}$. Unfortunately, a preliminary analysis  seems to suggest more physical input is needed in order to fix OPE coefficients from correlators involving $\hat{\Bm}_2$. More precisely, there are two different types of $4d$ multiplets that give rise to $2d$ operators of the same holomorphic dimension, and which we cannot disentangle without extra input.


\section{Consequences for the \texorpdfstring{$\Nm=2$}{N=2} landscape}
\label{sec:bounds}

In order to have a more comprehensive view of the landscape of $\Nm=2$ theories, we will supplement our new bound \eqref{eq:newbound} with the analytic bounds obtained in \cite{Beem:2013sza}. These were also obtained using the chiral algebra construction, but by looking only at the four-point function of affine Kac-Moody currents, instead of the full mixed system considered here. 
Among the most interesting bounds found in \cite{Beem:2013sza}, is the one associated to the vanishing of a $\hat{\Bm}_2$ multiplet in the singlet channel:
\be 
\frac{1}{k_{4d}} \leqslant \frac{12 c_{4d} + \text{dim}_G}{24 c_{4d} h^{\vee}}\, .
\label{eq:ckcbound}
\ee
Bounds of the form $k_{4d} \geqslant n(G)$ where $n(G)$ is a real number that only depends on the flavor symmetry were also obtained. With the exception of $G=\SU(2)$, all  of these bounds arise from the vanishing of a $\hat{\Bm}_2$ multiplet in a particular representation appearing in $\text{Sym}^2(\textbf{adj})$. We refer the reader to the original paper for the complete list.  
Here we will be mostly concerned with \eqref{eq:ckcbound} and \eqref{eq:newbound}.

Saturation of the analytic bounds implies relations among the different parameters appearing in the inequality.
An interesting question is whether it is possible to saturate \eqref{eq:ckcbound} and \eqref{eq:newbound} simultaneously, in the range of $k_{4d}$ allowed by the bounds of the type  $k_{4d} \geqslant n(G)$. It turns out this can happen for only a handful of flavor symmetries with specific values for the central charges $c_{4d}$ and $k_{4d}$. We list all the solutions in Tab.~\ref{Tab:Deligne}. It is worth pointing out that in this list the value of $k_{4d}$ saturates the bound of \cite{Beem:2013sza} for these groups (with the exception of the $\SU(2)$ case), namely $k_{4d} \geqslant \tfrac{h^\vee}{3}+2$. Remarkably, all the rank one theories of Tab.~\ref{Tab:rank1theories} are contained in this list. In addition, there are two solutions corresponding to flavor symmetries $\G_2$ and $\F_4$, which had already made an appearance in \cite{Beem:2013sza}, but whose associated four-dimensional SCFTs are not known to exist.
\begin{table}
\centering
\renewcommand{\arraystretch}{1.3}
\begin{tabular}{|c||c|c|c|c|c|c|c| c|}
\hline
$G$		&	$\SU(2)$	      & $\SU(3)$ 	       & $\G_2$ 	        & $\SO(8)$ 	  & $\F_4$ 	    & $\E_6$ 		    & $\E_7$ 		  & $\E_8$ 			\\ \hline\hline 
$c_{4d}$ 	    & $\frac{1}{2}$   & $\frac{2}{3}$   &    $\frac{5}{6}$    &  $\frac{7}{6}$ &$\frac{5}{3}$ & $\frac{13}{6}$  & $\frac{19}{6}$  & $\frac{31}{6}$  \\ \hline
$k_{4d}$ 	   & $\frac{8}{3}$    & $3$ 		       &    $\frac{10}{3}$ & $4$ 	  &	  $5$       & $6$ 			& $8$ 	      & $12$ 			\\ 
\hline
\end{tabular}
\caption{Simultaneous saturation of the new bound \eqref{eq:newbound} and the bound \eqref{eq:ckcbound} for the allowed range of $k_{4d}$.}
\label{Tab:Deligne}
\end{table}
In fact, this collection of algebras and central charges was also obtained in \cite{Beem:2013sza}, although for a given $k_{4d}$, the bounds there still allowed for smaller $c_{4d}$.
The pairs $(c_{4d},k_{4d})$ were only fixed after imposing that the Higgs branch of the theory corresponded to the one-instanton moduli space for $G$ instantons, \ie, by imposing Higgs branch relations that define the Joseph ideal (see \eg, \cite{Gaiotto:2008nz}). This is known to hold for the aforementioned rank one theories.
These relations imply the saturation of the bounds in all channels appearing in $\text{Sym}^2(\textbf{adj}) - \left(2 \textbf{adj}\right)$, where $\left(2\textbf{adj}\right)$ denotes the representation whose Dynkin labels are twice that of the adjoint. This requires $G$ to be given by the list in Tab.~\ref{Tab:Deligne} and, with the exception of $G=\SU(2)$, the values of $c_{4d}$ and $k_{4d}$ are fixed by the requirement that both \eqref{eq:ckcbound} and the $k_{4d} \geqslant n(G)$ bounds be saturated. 

From the two-dimensional perspective saturation of these bounds implies the presence of a null state in the chiral algebra. For \eqref{eq:ckcbound} this has as a consequence that the stress tensor of the $2d$ theory is  given by the Sugawara construction. Indeed, it is easy to check that solving for $c_{2d}$ gives the standard formula for the Sugawara central charge. This led \cite{Beem:2013sza} to  conjecture that the chiral algebras associated to these theories are affine Kac-Moody current algebras at level $k_{2d}=-\tfrac{k_{4d}}{2}$, where $k_{4d}$ is given in Tab.~\ref{Tab:Deligne}.\footnote{This was also studied in connection with the superconformal index in \cite{Buican:2015ina,Cordova:2015nma, Buican:2015tda}.}
On the other hand, saturation of \eqref{eq:newbound} signals the presence of a null state at level $h=4$. This state can be interpreted as a difference between two holomorphic operators, and its absence implies the $\SL(2)$ primary containing $:TT:$ can be written in terms of the $J$ current.

\begin{figure}[h!]
\centering
\includegraphics[scale=0.35]{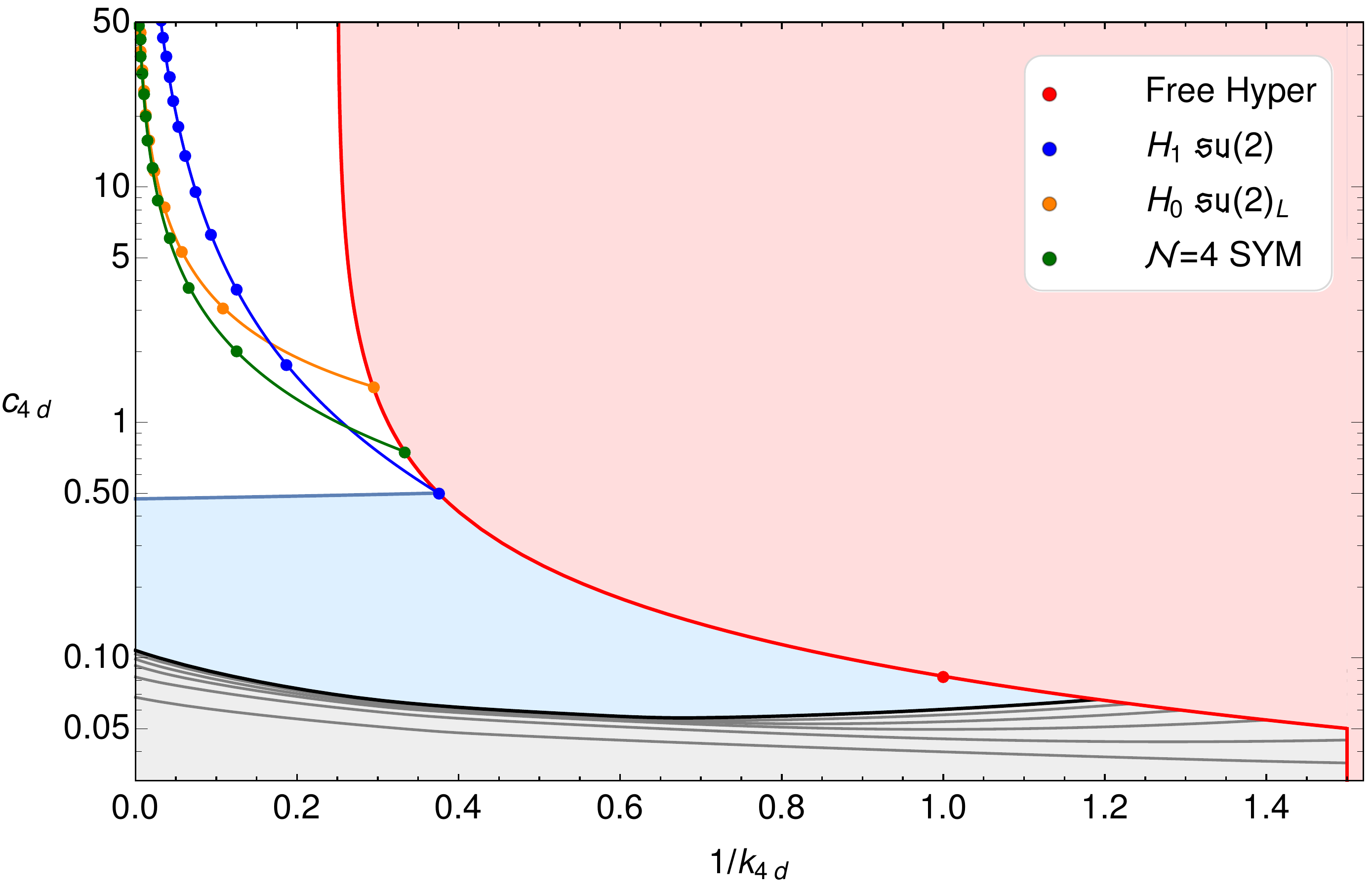}
\caption{Constraints on the $(c_{4d},k_{4d})$ plane for $\SU(2)$ flavor symmetries arising as a combination of the analytic bound \eqref{eq:newbound} (blue), the analytic bounds of \cite{Beem:2013sza} (red) and the numerical bound of \cite{Beem:2014zpa} (gray). We have marked the position of some known $\NN=2$ theories which contain an $\SU(2)$ flavor symmetry, namely the free hyper multiplet, $\Nm=4$ $SU(N)$ SYM, and the theories obtained by $N$ $D3$-branes probing an F-theory singularity of type $H_1$ (starting with the rank $N=1$ of Tab.~\ref{Tab:rank1theories}), and $H_0$ (for $N \geqslant2$ when it has an $\SU(2)$ flavor symmetry) \cite{Aharony:2007dj}.}
\label{fig:SU2landscape}
\end{figure}

Let us now analyze our new bound in the $(c_{4d},k_{4d})$ plane to have a clearer picture of how the $\Nm=2$ theory space looks.
In Fig.~\ref{fig:SU2landscape} we plot in blue the analytic bound \eqref{eq:newbound}, and in red the analytic bounds of \cite{Beem:2013sza} (\eqref{eq:ckcbound} and $k\geqslant n(G)$) for the simplest case of $G=\SU(2)$. We have added a sample of known theories with flavor group either equal to $\SU(2)$, or containing an $\SU(2)$ factor.
The value of $k_{4d}= n(G=\SU(2))=\frac{2}{3}$ is ruled out by \eqref{eq:ckcbound}, and now the smallest allowed $k_{4d}$ is $\frac{8}{3}$, for which $c_{4d}$ is fixed uniquely-- that of the $H_1$ theory. 
Note that the free hypermultiplet, which enjoys an $\SU(2)$ flavor symmetry, is ruled out by the analytic bounds. This is consistent since both bounds were obtained imposing the absence of higher spin currents. We have also superimposed the numerical bound obtained in \cite{Beem:2014zpa}. It is clear that our analytic bound is significantly stronger.\footnote{This numerical $\SU(2)$ bound had already been improved by the $c \geqslant \tfrac{11}{30}$ result of \cite{Liendo:2015ofa}, which applied to any local unitary SCFT, regardless of flavor symmetry.} The numerical bound exhibited a puzzling minimum that would have corresponded to an $\Nm=2$ theory with no conventional Coulomb branch, this now cannot be the case since that region is disallowed.
Finally, we note that the central charges of $\Nm=4$ SYM, which in $\Nm=2$ language have an $\SU(2)$ flavor symmetry, have low values when compared to the $\Nm=2$ families. Any improvement of our central charge bound will at best be saturated by $\Nm=4$, unless we restrict ourselves to pure $\Nm=2$ theories. This can be accomplished by imposing the absence of certain $\Dm$-type multiplets that contain the extra supercurrents of $\Nm=4$.

\begin{figure}[h!]
\begin{center}
\begin{subfigure}[b]{0.47\textwidth}
\includegraphics[scale=0.36]{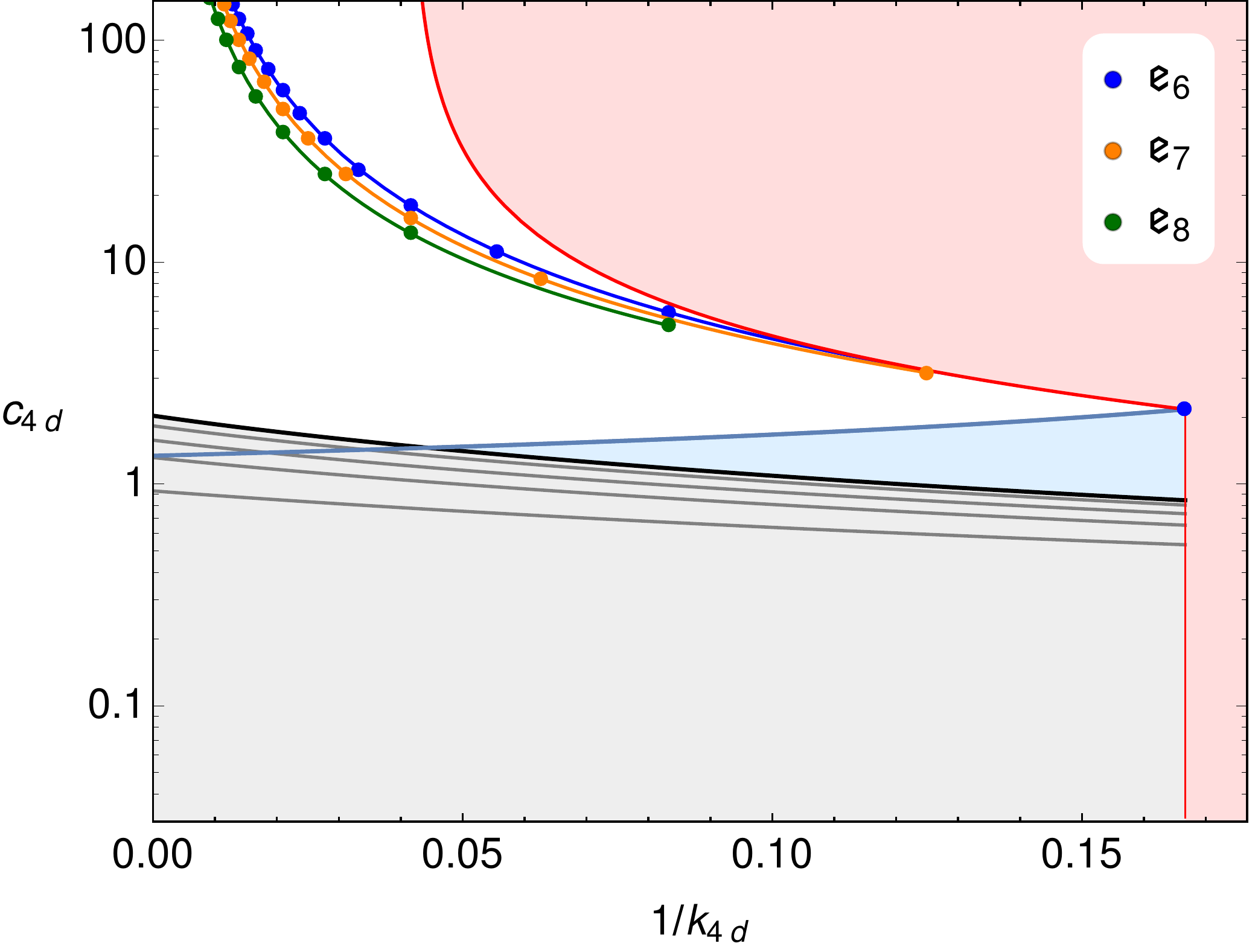}
\caption{$\E_6$}
\label{fig:E6landscape}
\end{subfigure}
\quad
\begin{subfigure}[b]{0.47\textwidth}
\includegraphics[scale=0.36]{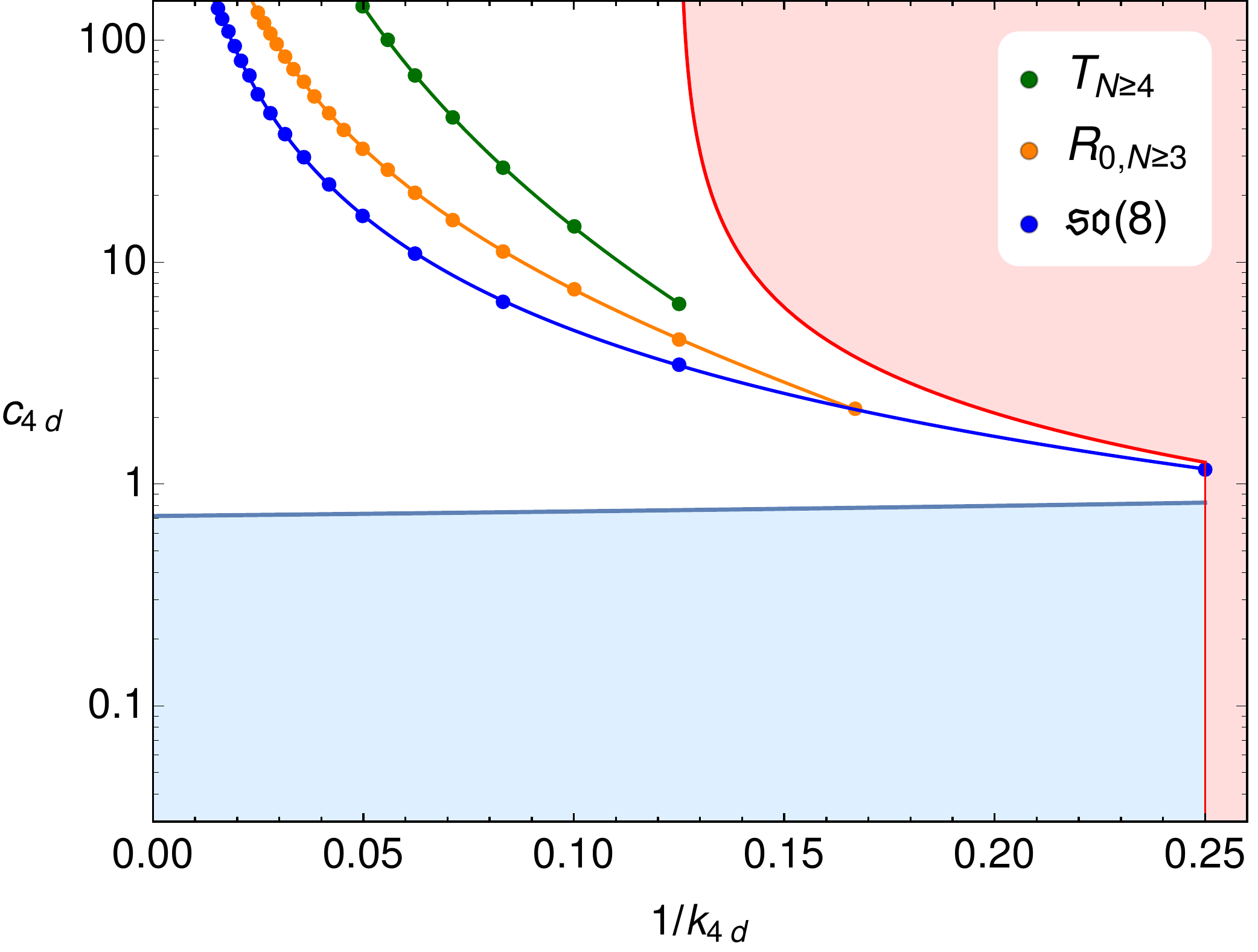}
\caption{$\SU(4)$}
\label{fig:SU4landscape}
\end{subfigure}
\end{center}
\caption{Constraints on the $(c_{4d},k_{4d})$ plane for two different flavor symmetries arising as a combination of the analytic bound \eqref{eq:newbound} (blue) and the analytic bounds of \cite{Beem:2013sza} (red). For the case of $\E_6$ the numerical bound of \cite{Beem:2014zpa} is also shown in gray. We also marked the position of several theories with at least $\E_6$ and $\SU(4)$ flavor symmetry, namely the rank $N$ version of some of the theories given in Tab.~\ref{Tab:rank1theories} \cite{Aharony:2007dj}, the $T_{N\geqslant4}$ family of theories of \cite{Gaiotto:2009we} which have an $\SU(N)^3$ flavor symmetry, and the $R_{0,N\geqslant3}$ of \cite{Chacaltana:2010ks} which have an $\SU(2) \times \SU(2N)$ flavor symmetry (for $N=3$ the flavor symmetry enhances and it corresponds to the rank one $\E_6$).}
\label{fig:E6SU4landscape}
\end{figure}

As a second example we turn to $G=\E_6$ in Fig.~\ref{fig:E6landscape}. As in the $\SU(2)$ case there is an intersection between the analytic bounds \eqref{eq:newbound} (shown in blue) and \eqref{eq:ckcbound} (shown in red), but now the intersection point also coincides with the minimal allowed value of $k_{4d}$ (also in red). This plot is a representative of what happens for all the remaining groups in Tab.~\ref{Tab:Deligne}, in which the values of $(c_{4d},k_{4d})$ are selected by a triple intersection of bounds, $\SU(2)$ being the only exception in which the $k_{4d}$ bound is a bit weaker.
The $\E_6$ plot can also be complemented with the numerical bounds of \cite{Beem:2014zpa}. Contrary to the $\SU(2)$ case, the numerical bounds allow to carve out a larger portion of the $(c_{4d},k_{4d})$ plane. This is good news, the analytic bound is strong for low values of $k_{4d}$, but for larger values the numerical bound does better. The strongest results are achieved by complementing analytic and numerical techniques. In this plot we also marked the position of a few theories with flavor group equal to, or containing $\E_6$ to give an idea of how the $\E_6$ landscape looks.

Finally, as a representative of all the other flavor groups for which there is no intersection of \eqref{eq:newbound} and \eqref{eq:ckcbound}, we show the case of $\SU(4)$ in Fig.~\ref{fig:SU4landscape}, where the first bound is shown in blue and the second in red. The gap on the right side means that for the smallest possible $k_{4d}$ there is a range of allowed values for the central charge. Indeed, it is not hard to find a theory which lies inside this range. The rightmost dot in the figure represents $\SU(2)$ superconformal QCD, which has an enhanced $\SO(8)$ symmetry and corresponds to the $\SO(8)$ entry of Tab.~\ref{Tab:Deligne}. Here we are considering it by embedding an $\SU(4)$ factor inside the $\SO(8)$. For larger rank theories, either $\SU(N)$, $\SO(N)$, $\USp(N)$ the gap becomes larger and the bounds are less constraining. As $N$ increases, the upper and lower bounds get more separated, with the difference between them growing like $N^2$.

To conclude, let us discuss how one could improve on the results presented here. Looking at the landscape plots, it is natural to ask whether a bound of the form $\tfrac{c_{4d}}{k_{4d}} \geqslant \ldots$ could ever be found from the bootstrap.
Holographic arguments suggest that such a bound should exist, at least for theories with a weakly coupled gravitational description.
On the dual $AdS$, the central charges $c_{4d}$ and $k_{4d}$ are related to the strength of the gravitational and gauge couplings, and the statement that $\tfrac{c_{4d}}{k_{4d}}$ is bounded from below would follow from the statement that \emph{gravity is the weakest force} \cite{ArkaniHamed:2006dz}.\footnote{See for example \cite{Nakayama:2015hga} for an $AdS$ discussion of this statement and its implications for CFTs.} The fact that the numerical $\E_6$ bound improves on the analytic bound for large $k$, seems to hint that this information is indeed coded in the $\Nm=2$ crossing symmetry equations. We hope to come back to this problem in the future.


\acknowledgments

We have greatly benefited from discussions with
L. Rastelli and
V. Schomerus.
The research leading to these results has received funding from the People Programme (Marie Curie Actions) of the European Union’s Seventh Framework Programme FP7/2007-2013/ under REA Grant Agreement No 317089 (GATIS).
P. L. is supported by SFB 647 ``Raum-Zeit-Materie. Analytische und Geometrische Strukturen''.

\bibliography{./aux/biblio}

\providecommand{\href}[2]{#2}\begingroup\raggedright\begin{thebibliography}{10}

\bibitem{Beem:2013sza}
C.~Beem, M.~Lemos, P.~Liendo, W.~Peelaers, L.~Rastelli, and B.~C. van Rees,
  {\it {Infinite Chiral Symmetry in Four Dimensions}},  {\em Communications in
  Mathematical Physics} {\bf 336} (2015), no.~3 1359--1433,
  [\href{http://arxiv.org/abs/1312.5344}{{\tt arXiv:1312.5344}}].

\bibitem{Beem:2014rza}
C.~Beem, W.~Peelaers, L.~Rastelli, and B.~C. van Rees, {\it {Chiral algebras of
  class S}},  {\em JHEP} {\bf 1505} (2015) 020,
  [\href{http://arxiv.org/abs/1408.6522}{{\tt arXiv:1408.6522}}].

\bibitem{Lemos:2014lua}
M.~Lemos and W.~Peelaers, {\it {Chiral Algebras for Trinion Theories}},  {\em
  JHEP} {\bf 1502} (2015) 113, [\href{http://arxiv.org/abs/1411.3252}{{\tt
  arXiv:1411.3252}}].

\bibitem{Liendo:2015ofa}
P.~Liendo, I.~Ramirez, and J.~Seo, {\it {Stress-tensor OPE in N=2
  Superconformal Theories}},  \href{http://arxiv.org/abs/1509.00033}{{\tt
  arXiv:1509.00033}}.

\bibitem{Chacaltana:2014nya}
O.~Chacaltana, J.~Distler, and A.~Trimm, {\it {A Family of $4D$ $\mathcal{N}=2$
  Interacting SCFTs from the Twisted $A_{2N}$ Series}},
  \href{http://arxiv.org/abs/1412.8129}{{\tt arXiv:1412.8129}}.

\bibitem{Buican:2015ina}
M.~Buican and T.~Nishinaka, {\it {On the Superconformal Index of
  Argyres-Douglas Theories}},  \href{http://arxiv.org/abs/1505.05884}{{\tt
  arXiv:1505.05884}}.

\bibitem{Cordova:2015nma}
C.~Cordova and S.-H. Shao, {\it {Schur Indices, BPS Particles, and
  Argyres-Douglas Theories}},  \href{http://arxiv.org/abs/1506.00265}{{\tt
  arXiv:1506.00265}}.

\bibitem{Buican:2015tda}
M.~Buican and T.~Nishinaka, {\it {Argyres-Douglas Theories, the Macdonald
  Index, and an RG Inequality}},  \href{http://arxiv.org/abs/1509.05402}{{\tt
  arXiv:1509.05402}}.

\bibitem{Song:2015wta}
J.~Song, {\it {Superconformal indices of generalized Argyres-Douglas theories
  from 2d TQFT}},  \href{http://arxiv.org/abs/1509.06730}{{\tt
  arXiv:1509.06730}}.

\bibitem{Cecotti:2015lab}
S.~Cecotti, J.~Song, C.~Vafa, and W.~Yan, {\it {Superconformal Index, BPS
  Monodromy and Chiral Algebras}},  \href{http://arxiv.org/abs/1511.01516}{{\tt
  arXiv:1511.01516}}.

\bibitem{Beem:2014kka}
C.~Beem, L.~Rastelli, and B.~C. van Rees, {\it {$ \mathcal{W} $ symmetry in six
  dimensions}},  {\em JHEP} {\bf 1505} (2015) 017,
  [\href{http://arxiv.org/abs/1404.1079}{{\tt arXiv:1404.1079}}].

\bibitem{Chester:2014mea}
S.~M. Chester, J.~Lee, S.~S. Pufu, and R.~Yacoby, {\it {Exact Correlators of
  BPS Operators from the 3d Superconformal Bootstrap}},  {\em JHEP} {\bf 03}
  (2015) 130, [\href{http://arxiv.org/abs/1412.0334}{{\tt arXiv:1412.0334}}].

\bibitem{Beem:2013qxa}
C.~Beem, L.~Rastelli, and B.~C. van Rees, {\it {The N=4 Superconformal
  Bootstrap}},  {\em Phys.Rev.Lett.} {\bf 111} (2013) 071601,
  [\href{http://arxiv.org/abs/1304.1803}{{\tt arXiv:1304.1803}}].

\bibitem{Argyres:1995xn}
P.~C. Argyres, M.~R. Plesser, N.~Seiberg, and E.~Witten, {\it {New N=2
  superconformal field theories in four-dimensions}},  {\em Nucl. Phys.} {\bf
  B461} (1996) 71--84, [\href{http://arxiv.org/abs/hep-th/9511154}{{\tt
  hep-th/9511154}}].

\bibitem{Minahan:1996fg}
J.~A. Minahan and D.~Nemeschansky, {\it {An N=2 superconformal fixed point with
  E(6) global symmetry}},  {\em Nucl.Phys.} {\bf B482} (1996) 142--152,
  [\href{http://arxiv.org/abs/hep-th/9608047}{{\tt hep-th/9608047}}].

\bibitem{Minahan:1996cj}
J.~A. Minahan and D.~Nemeschansky, {\it {Superconformal fixed points with E(n)
  global symmetry}},  {\em Nucl.Phys.} {\bf B489} (1997) 24--46,
  [\href{http://arxiv.org/abs/hep-th/9610076}{{\tt hep-th/9610076}}].

\bibitem{Cheung:1997id}
Y.-K.~E. Cheung, O.~J. Ganor, and M.~Krogh, {\it {Correlators of the global
  symmetry currents of 4-D and 6-D superconformal theories}},  {\em Nucl.Phys.}
  {\bf B523} (1998) 171--192, [\href{http://arxiv.org/abs/hep-th/9710053}{{\tt
  hep-th/9710053}}].

\bibitem{Argyres:2007cn}
P.~C. Argyres and N.~Seiberg, {\it {S-duality in N=2 supersymmetric gauge
  theories}},  {\em JHEP} {\bf 0712} (2007) 088,
  [\href{http://arxiv.org/abs/0711.0054}{{\tt arXiv:0711.0054}}].

\bibitem{Aharony:2007dj}
O.~Aharony and Y.~Tachikawa, {\it {A Holographic computation of the central
  charges of d=4, N=2 SCFTs}},  {\em JHEP} {\bf 0801} (2008) 037,
  [\href{http://arxiv.org/abs/0711.4532}{{\tt arXiv:0711.4532}}].

\bibitem{Sen:1996vd}
A.~Sen, {\it {F theory and orientifolds}},  {\em Nucl.Phys.} {\bf B475} (1996)
  562--578, [\href{http://arxiv.org/abs/hep-th/9605150}{{\tt hep-th/9605150}}].

\bibitem{Banks:1996nj}
T.~Banks, M.~R. Douglas, and N.~Seiberg, {\it {Probing F theory with branes}},
  {\em Phys.Lett.} {\bf B387} (1996) 278--281,
  [\href{http://arxiv.org/abs/hep-th/9605199}{{\tt hep-th/9605199}}].

\bibitem{Dasgupta:1996ij}
K.~Dasgupta and S.~Mukhi, {\it {F theory at constant coupling}},  {\em
  Phys.Lett.} {\bf B385} (1996) 125--131,
  [\href{http://arxiv.org/abs/hep-th/9606044}{{\tt hep-th/9606044}}].

\bibitem{Aharony:1998xz}
O.~Aharony, A.~Fayyazuddin, and J.~M. Maldacena, {\it {The Large N limit of
  N=2, N=1 field theories from three-branes in F theory}},  {\em JHEP} {\bf
  9807} (1998) 013, [\href{http://arxiv.org/abs/hep-th/9806159}{{\tt
  hep-th/9806159}}].

\bibitem{Gaiotto:2008nz}
D.~Gaiotto, A.~Neitzke, and Y.~Tachikawa, {\it {Argyres-Seiberg duality and the
  Higgs branch}},  {\em Commun.Math.Phys.} {\bf 294} (2010) 389--410,
  [\href{http://arxiv.org/abs/0810.4541}{{\tt arXiv:0810.4541}}].

\bibitem{Beem:2014zpa}
C.~Beem, M.~Lemos, P.~Liendo, L.~Rastelli, and B.~C. van Rees, {\it {The
  ${\mathcal N}=2$ superconformal bootstrap}},
  \href{http://arxiv.org/abs/1412.7541}{{\tt arXiv:1412.7541}}.

\bibitem{Lemos:2015awa}
M.~Lemos and P.~Liendo, {\it {Bootstrapping ${\mathcal N}=2$ chiral
  correlators}},  \href{http://arxiv.org/abs/1510.03866}{{\tt
  arXiv:1510.03866}}.

\bibitem{Dolan:2001tt}
F.~Dolan and H.~Osborn, {\it {Superconformal symmetry, correlation functions
  and the operator product expansion}},  {\em Nucl.Phys.} {\bf B629} (2002)
  3--73, [\href{http://arxiv.org/abs/hep-th/0112251}{{\tt hep-th/0112251}}].

\bibitem{Nirschl:2004pa}
M.~Nirschl and H.~Osborn, {\it {Superconformal Ward identities and their
  solution}},  {\em Nucl.Phys.} {\bf B711} (2005) 409--479,
  [\href{http://arxiv.org/abs/hep-th/0407060}{{\tt hep-th/0407060}}].

\bibitem{Dolan:2004mu}
F.~A. Dolan, L.~Gallot, and E.~Sokatchev, {\it {On four-point functions of
  1/2-BPS operators in general dimensions}},  {\em JHEP} {\bf 09} (2004) 056,
  [\href{http://arxiv.org/abs/hep-th/0405180}{{\tt hep-th/0405180}}].

\bibitem{Kinney:2005ej}
J.~Kinney, J.~M. Maldacena, S.~Minwalla, and S.~Raju, {\it {An Index for 4
  dimensional super conformal theories}},  {\em Commun.Math.Phys.} {\bf 275}
  (2007) 209--254, [\href{http://arxiv.org/abs/hep-th/0510251}{{\tt
  hep-th/0510251}}].

\bibitem{Gadde:2011ik}
A.~Gadde, L.~Rastelli, S.~S. Razamat, and W.~Yan, {\it {The 4d Superconformal
  Index from q-deformed 2d Yang-Mills}},  {\em Phys.Rev.Lett.} {\bf 106} (2011)
  241602, [\href{http://arxiv.org/abs/1104.3850}{{\tt arXiv:1104.3850}}].

\bibitem{Gadde:2011uv}
A.~Gadde, L.~Rastelli, S.~S. Razamat, and W.~Yan, {\it {Gauge Theories and
  Macdonald Polynomials}},  {\em Commun.Math.Phys.} {\bf 319} (2013) 147--193,
  [\href{http://arxiv.org/abs/1110.3740}{{\tt arXiv:1110.3740}}].

\bibitem{Dolan:2002zh}
F.~Dolan and H.~Osborn, {\it {On short and semi-short representations for
  four-dimensional superconformal symmetry}},  {\em Annals Phys.} {\bf 307}
  (2003) 41--89, [\href{http://arxiv.org/abs/hep-th/0209056}{{\tt
  hep-th/0209056}}].

\bibitem{Arutyunov:2001qw}
G.~Arutyunov, B.~Eden, and E.~Sokatchev, {\it {On nonrenormalization and OPE in
  superconformal field theories}},  {\em Nucl.Phys.} {\bf B619} (2001)
  359--372, [\href{http://arxiv.org/abs/hep-th/0105254}{{\tt hep-th/0105254}}].

\bibitem{Israel}
I.~Ramirez. \emph{Work in progress}.

\bibitem{Maldacena:2011jn}
J.~Maldacena and A.~Zhiboedov, {\it {Constraining Conformal Field Theories with
  A Higher Spin Symmetry}},  {\em J.Phys.} {\bf A46} (2013) 214011,
  [\href{http://arxiv.org/abs/1112.1016}{{\tt arXiv:1112.1016}}].

\bibitem{Alba:2013yda}
V.~Alba and K.~Diab, {\it {Constraining conformal field theories with a higher
  spin symmetry in d=4}},  \href{http://arxiv.org/abs/1307.8092}{{\tt
  arXiv:1307.8092}}.

\bibitem{Belavin:1984vu}
A.~A. Belavin, A.~M. Polyakov, and A.~B. Zamolodchikov, {\it {Infinite
  Conformal Symmetry in Two-Dimensional Quantum Field Theory}},  {\em Nucl.
  Phys.} {\bf B241} (1984) 333--380.

\bibitem{Gaiotto:2009we}
D.~Gaiotto, {\it {N=2 dualities}},  {\em JHEP} {\bf 1208} (2012) 034,
  [\href{http://arxiv.org/abs/0904.2715}{{\tt arXiv:0904.2715}}].

\bibitem{Chacaltana:2010ks}
O.~Chacaltana and J.~Distler, {\it {Tinkertoys for Gaiotto Duality}},  {\em
  JHEP} {\bf 1011} (2010) 099, [\href{http://arxiv.org/abs/1008.5203}{{\tt
  arXiv:1008.5203}}].

\bibitem{ArkaniHamed:2006dz}
N.~Arkani-Hamed, L.~Motl, A.~Nicolis, and C.~Vafa, {\it {The String landscape,
  black holes and gravity as the weakest force}},  {\em JHEP} {\bf 06} (2007)
  060, [\href{http://arxiv.org/abs/hep-th/0601001}{{\tt hep-th/0601001}}].

\bibitem{Nakayama:2015hga}
Y.~Nakayama and Y.~Nomura, {\it {Weak Gravity Conjecture in AdS/CFT}},
  \href{http://arxiv.org/abs/1509.01647}{{\tt arXiv:1509.01647}}.

\end{thebibliography}\endgroup
\bibliographystyle{./aux/JHEP}

\end{document}